\documentclass[10pt]{iopart}
\usepackage{epsfig}
\begin{document}
\title[Gravitational waves from extreme mass ratio inspirals]
{Gravitational waves from extreme mass ratio inspirals:
Challenges in mapping the spacetime of massive, compact objects}
\author{Scott A Hughes\footnote[1]{E-mail: hughes@tapir.caltech.edu}}
\address{
Theoretical Astrophysics, California Institute of Technology,
Pasadena, CA 91125}
\begin{abstract}
In its final year of inspiral, a stellar mass ($1 - 10 M_\odot$) body
orbits a massive ($10^5 - 10^7 M_\odot$) compact object about $10^5$
times, spiralling from several Schwarzschild radii to the last stable
orbit.  These orbits are deep in the massive object's strong field, so
the gravitational waves that they produce probe the strong field
nature of the object's spacetime.  Measuring these waves can, in
principle, be used to ``map'' this spacetime, allowing observers to
test whether the object is a black hole or something more exotic.
Such measurements will require a good theoretical understanding of
wave generation during inspiral.  In this article, I discuss the major
theoretical challenges standing in the way of building such maps from
gravitational-wave observations, as well as recent progress in
producing extreme mass ratio inspirals and waveforms.
\end{abstract}

\pacs{04.25.Nx, 04.30.-w, 04.30.Nk}

\vskip 2pc

Chandrasekhar has described black holes as ``the most perfect
macroscopic objects there are in the universe'' {\cite{chandra83}}.
This description refers to their simplicity, depending (in
astrophysical contexts) solely on the mass and spin of the hole.  This
dependence, in turn, follows from the black hole uniqueness theorems
{\cite{israel,carter,robinson,price1,price2}}, which guarantee that
all of the other ``hairs'' will radiate away during the hole's
formation.  If general relativity correctly describes gravity, then
the massive compact objects at the centers of most galaxies are
probably described exactly by the Kerr rotating black hole metric.  In
principle, this description can be tested using LISA {\cite{lisa}}:
gravitational-wave observations of ``small'' ($1 - 10\,M_\odot$)
bodies spiralling into ``large'' ($10^5 - 10^7\,M_\odot$) compact
objects can be used to map the spacetime of the large compact object,
testing whether it is a Kerr black hole or some other exotic object.
In this article, I discuss the theoretical challenges that must be met
before such maps can be constructed, as well as recent progress.

Fintan Ryan {\cite{fintanmeasure}} first showed that a body's
spacetime can be mapped with gravitational waves.  The spacetime of a
massive object arises from its multipole moment structure.  These
multipole moments come in two varieties, mass ($M_l$) and current
($S_l$).  If $\rho(\vec r)$ is the mass density at position $\vec r$,
and $\vec v(\vec r)$ is the fluid velocity at $\vec r$, then these
multipoles are roughly
\begin{equation}
M_l \simeq \int \rmd^3r\,r^l\rho({\vec r})\,,
\qquad S_l \simeq \int \rmd^3r\,r^{l-1}\left[{\vec r}\times
{\vec v(\vec r)}
\rho(\vec r)\right]
\label{multipoledefs}
\end{equation}
For a black hole, the mass and current multipole moments take a far
simpler form:
\begin{equation}
M^{\rm BH}_l + \rmi S^{\rm BH}_l = M(\rmi a)^l
\label{BHmultipoles}
\end{equation}
Because these moments directly determine the spacetime, the orbits of
and radiation emitted by a small\footnote{``Small'' means that the
orbiting body does not significantly change the spacetime.} body
moving in this spacetime are strongly influenced by the massive
object's multipoles.  Measuring the radiation allows one (in
principle) to measure the multipolar structure of the massive object.

Much work remains before it will be possible to measure an exotic
body's multipole moments in practice.  In particular, techniques must
be developed to solve the wave equation for gravitational radiation
from orbits of exotic objects.  This is not a terribly difficult
matter for black hole spacetimes --- following Teukolsky's 1972
discovery {\cite{teuk72}} that the wave equation for radiation
propagating in Kerr spacetimes is separable, an array of calculational
technology has been developed for studying the generation and
propagation of such radiation (see, {\it e.g.}, {\cite{hughes_gsn}}
for review and discussion).  Very little work has been done for
radiation generated in and propagating through the spacetime of more
exotic objects: Ryan {\cite{fintanscalar}} has examined the scalar
waves produced by highly constrained orbits (circular and equatorial)
of objects with arbitrary multipole moments.  We are a long way from
understanding the waveforms created by orbits of non-black hole
objects.  Such an understanding will be needed --- at least to some
degree --- in order to probe the multipole character of massive
compact bodies.

For now, we simplify the problem by focusing upon waves generated when
the massive object is a black hole.  Of greatest interest is
understanding the waves emitted as a small, spinning body spirals into
a massive Kerr black hole.  Neglecting radiation reaction, the motion
of such a body is governed by the Papapetrou equations {\cite{dixon}}:
\begin{eqnarray}
{Dp^\mu\over D\tau} &= -{1\over2} R^\mu_{\nu\rho\sigma} v^\nu
S^{\rho\sigma}\nonumber\\
{DS^{\mu\nu}\over D\tau} &= p^\mu v^\nu - p^\nu v^\mu
\label{papapetrou}
\end{eqnarray}
In these equations, $D/D\tau$ denotes a covariant derivative along the
small body's trajectory, $R^\mu_{\nu\rho\sigma}$ is the Riemann
curvature tensor of the background spacetime, $S^{\mu\nu}$ is a tensor
related to the spin of the body, $v^\mu = dz^\mu/d\tau$ where
$z^\mu(\tau)$ is the coordinate worldline of the body, and $p^\mu$ is
a generalization of the body's 4-momentum that incorporates spin.
These equations show that as the small body orbits, its spin couples
to the curvature of the massive black hole.  In the weak field, this
coupling is known to lead to precessional effects which modulate the
phase and amplitude of the gravitational waveform
{\cite{precmodulation}}.  In the strong field, the coupling might
become extremely important.  It is likely that, at least for certain
parameter values, the orbital character will become chaotic.  Janna
Levin {\cite{janna2000}} has shown that integrating the second
post-Newtonian equivalents of {\Eref{papapetrou}} leads to chaotic
motion: orbits with similar initial conditions evolve to vastly
different configurations.  This could make data analysis extremely
difficult --- matched filtering, for example, would require an
enormous number of templates in order to cover all possible inspirals.
When radiation reaction is included, the effects of chaos in Levin's
work become less extreme: she focused upon sources of interest to
ground-based detectors, and the number of orbits visible to those
detectors is not very large.  With LISA, though, we expect to see
around $10^5$ orbits.  Chaotic evolution could lead to a dramatic
divergence of outcomes from similar initial conditions, rendering data
analysis practically impossible.  Detailed studies of these orbits
with parameters relevant to LISA are urgently needed.

Since spin opens an as-yet-poorly-understood can of worms, we will
ignore it for now, treating the small body as a point perturbation to
the Kerr spacetime.  At zeroth order, this small body moves on a Kerr
geodesic.  For such orbits, a relatively mature radiation reaction
formalism (see, {\it e.g.}, {\cite{hughes_rr}} and references therein)
has been developed that finds the first order radiative corrections to
this motion, allowing one to compute the trajectory that the body
follows as it spirals into the black hole, and the waveforms that it
generates.  This formalism uses the zeroth order geodesic motion as a
source for the first order corrections.  As such, it assumes that the
inspiral is adiabatic: the timescale for radiation reaction to change
the orbit's characteristics is much smaller than the orbit's dynamical
timescale.

For the most interesting orbits --- eccentric, inclined orbits in the
strong field of rotating black holes --- it is not clear at the
present time if this adiabatic approximation is reasonable.  It is
somewhat difficult to define the orbit's dynamical timescale in this
interesting case.  The issue is that there are three physically
meaningful timescales: the time $T_\phi$ for the body to cover $0 \le
\phi \le 2 \pi$; the time $T_r$ for the body to move from $r_{\rm
max}$ to $r_{\rm min}$ and back; and the time $T_\theta$ for the body
to move from its highest latitude $\theta_{\rm min}$ to its lowest
$\theta_{\rm max}$ and back.  These timescales generically are rather
different.  When the orbit is constrained (eccentric but equatorial,
or inclined but circular) only two of these timescales are non-zero.
One can analyze the orbit in a frame that rotates at frequency $\Omega
= 2\pi/T_\phi$, cancelling out the $\phi$ motion.  It is then simple
to define the orbit's dynamical timescale: it is $T_r$ (for eccentric,
equatorial orbits) or $T_\theta$ (for inclined, circular orbits) (see
{\cite{hughes_rr}} for more detail).

\begin{figure}
\begin{center}
\epsfig{file = 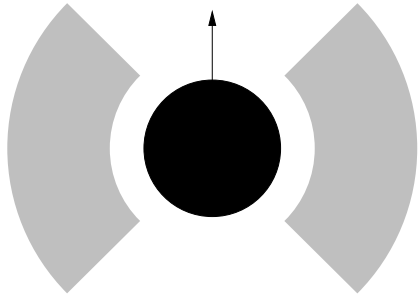, width = 5.5cm}
\end{center}
\caption{A slice of the orbital torus.  The dark circle is the central
Kerr black hole; the grey region is the volume which is filled by the
orbiting body.  The full torus is given by rotating this figure about
the spin axis (indicated by the arrow).}
\label{torus}
\end{figure}

When the motion is not constrained, this trick does not work.  To
understand the behavior of the orbit for the general case, consider
{\Fref{torus}}.  This figure shows a 2-dimensional slice of the volume
that is filled by a small body as it orbits the black hole: as it
moves from $r_{\rm min}$ to $r_{\rm max}$ it simultaneously moves
between $\theta_{\rm min}$ and $\theta_{\rm max}$.  Rotating this
figure about the black hole's spin axis, we see that the orbit
ergodically fills a torus in the spacetime near the hole's horizon.
For adiabaticity to be a good approximation, the orbiting body must
come ``close'' to every point in this volume.  (The meaning of
``close'' is of course rather ambiguous.  ``Close enough'' depends
upon the accuracy that one requires, which in turn depends upon how
well one needs to know the phase of the waveform.)  An important
problem is to determine how long it takes for an orbiting body to come
``close enough'' to all points in this volume for parameters that are
of interest to LISA observations (strong field, large black hole spin,
high eccentricity, and arbitary inclination angle).  If this time
turns out to be larger than the radiation reaction timescale then
further studies of these waveforms will require radiation reaction
forces that do not rely on adiabaticity
{\cite{quinnwald,mst,wiseman,ori,barackori,burko,lousto}}.  The phase
evolution of waveforms that one constructs in these cases may have a
strong dependence on the orbiting body's initial conditions --- they
may be ``effectively chaotic'' in the words of Schutz
{\cite{schutz_eff_chaos}}.

Even if it turns out that the issue of adiabaticity is not a serious
problem, there is still an important challenge that must be faced in
order to evolve generic Kerr orbits.  Such orbits are characterized by
three conserved quantities: the energy $E$, the $z$-component of
angular momentum $L_z$, and the Carter constant $Q$.  For
Schwarzschild black holes, $Q$ is just $L_x^2 + L_y^2$, the other
angular momentum components.  Interpretation of the Carter constant
becomes less clear as $a$ increases (the geometry becomes oblate,
confusing the meaning of $x$, $y$, and $z$, and frame dragging
entangles the $t$ and $\phi$ coordinates), but it is useful to regard
it essentially as the ``rest'' of the small body's angular momentum.
At present, the most mature computational formalisms (such as that
described in {\cite{hughes_rr}}) cannot evolve $Q$.  These formalisms
work by a method of ``flux-balancing'': from the flux of gravitational
waves going to infinity and down the hole's event horizon, one can
easily deduce the change in $E$ and $L_z$ because of
gravitational-wave emission.  One cannot easily deduce the change in
$Q$, except in special cases (for eccentric equatorial orbits, $Q = 0$
at all times; for inclined, circular orbits theorems which prove that
the orbit adiabatically remains circular
{\cite{kennefickori,fintancircular,mino}} allow one to express $dQ/dt$
as a function of $dE/dt$ and $dL_z/dt$).  Although clever methods may
make it possible to evolve $Q$ just by examining the radiation flux at
infinity and at the horizon (see Wolfgang Tichy's contribution to
these proceedings), it may turn out that a local radiation reaction
force will be needed.

At this point, the current unsolved or poorly understood issues have
narrowed the class of sources that are well understood rather
severely.  Current computational technology is limited to
understanding the adiabatic evolution of spinless bodies on
constrained orbits --- either eccentric equatorial orbits or inclined
circular orbits.  For the remainder of this article, I will focus on
circular inclined orbits, as described in {\cite{hughes_rr}}; Daniel
Kennefick and Kostas Glampedakis are developing an analysis of
eccentric equatorial orbits using a similar formalism.

\begin{figure}
\begin{center}
\epsfig{file = 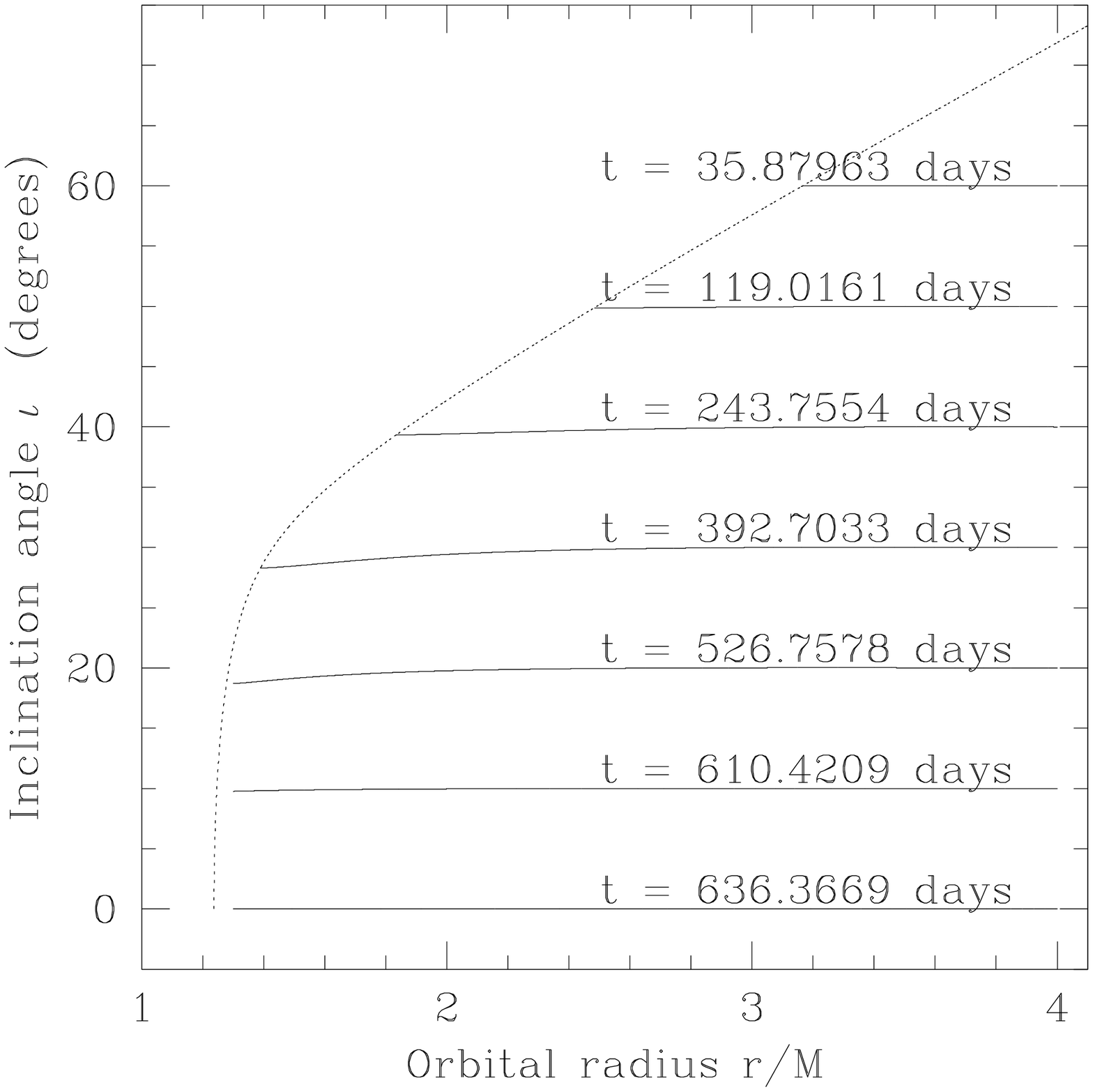, width = 6.485cm}
\epsfig{file = 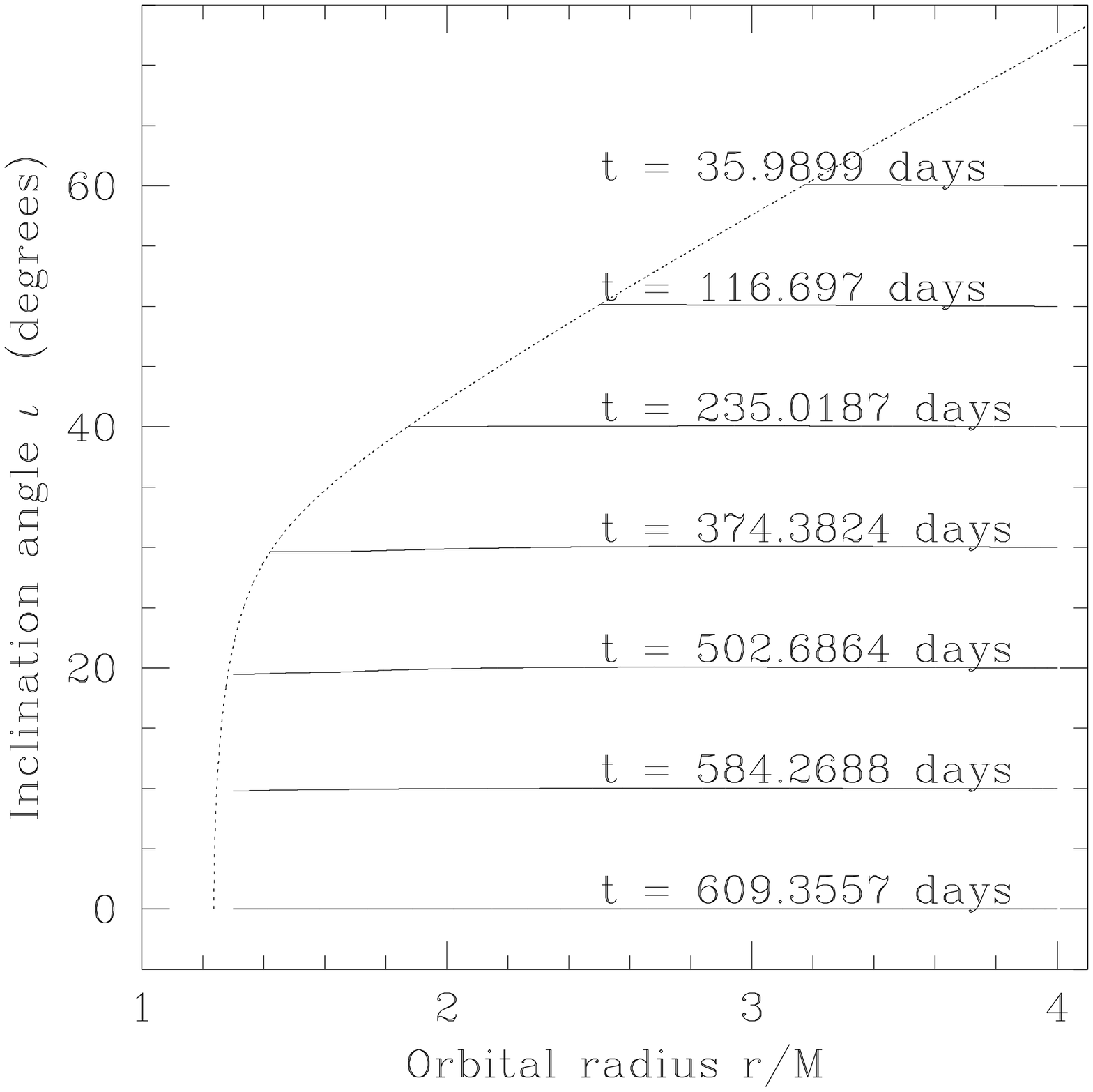, width = 6.485cm}
\end{center}
\caption{Inspiral from $r = 4 M$ to the last stable orbit (indicated
by the dotted line).  The massive black hole has $M = 10^6\, M_\odot$
and $a = 0.998 M$; the small body has $\mu = 1\,M_\odot$.  The
left-hand panel shows the inspiral including both flux to infinity and
flux down the horizon; the right-hand panel includes only flux to
infinity.  Notice that the horizon flux typically {\it slows} the
inspiral.}
\label{sequence}
\end{figure}

Using the formalism and code described in {\cite{hughes_rr}}, I have
studied the inspirals and associated gravitational waveforms for a
large number of strong-field initial conditions.  The results for $a =
0.998M$ are shown in the left-hand panel of {\Fref{sequence}}.  This
figure shows the inclination angle $\iota$ of the orbit as the small
body spirals from $r = 4 M$ to the last stable orbit (LSO); it also
shows the number of days that pass along this sequence.  In all cases
the trajectory is nearly flat --- the inclination does not change very
much as the small body inspirals.  The inclination evolution is even
flatter at smaller values of $a$ {\cite{hughes_inprep}}.  Curt Cutler
has suggested that this might be used as the basis of an approximate
scheme for evolving the Carter constant for generic orbits: if it is
true in general that the inclination angle does not change very
strongly, then it might be a reasonable approximation to set the
change to zero.  This condition would constrain the evolution of $Q$.
Cutler's approximation may be useful for developing approximate
waveforms for the study of data analysis tools.

Turn now to the right-hand panel of {\Fref{sequence}}.  This panel is
identical to the left-hand panel except that the flux of radiation
down the massive black hole's event horizon has been ignored in
constructing the inspiral trajectory.  Although the shape does not
change very much, the time it takes to inspiral is significantly {\it
smaller}: the horizon flux {\it slows} the inspiral by several weeks
at low inclination angle.  This is a significant effect --- change of
the inspiral time by such a large amount should be easily measurable.
At first glance, it is also rather counterintuitive: one expects the
hole's event horizon to be a sink of energy, so the inspiral would
speed up as radiation flows into the hole.  This simple picture is
wrong when the hole rotates.  A more accurate picture can be developed
by considering the tidal coupling of the black hole to the
inspiralling body.  The tidal field of the small body will distort the
hole, raising ``bulges'' in the event horizon {\cite{hartle}}.  These
bulges exert a torque back on the small body.  When the hole is
rapidly spinning, the bulges are dragged ahead of the orbiting body so
that this torque tends to increase the orbiting particle's energy.
This partially offsets the energy that is lost from radiation to
infinity, slowing the inspiral.

One of the major goals of this analysis is to produce gravitational
waveforms.  Using the inspiral trajectories shown in the left-hand
panel of {\Fref{sequence}}, I have developed the associated
gravitational waveforms:
\begin{equation}
\fl
h_+(t) - \rmi h_\times(t) = {\mu\over D}\sum_{l = 2}^\infty
\sum_{m = -l}^l\sum_{k = -\infty}^\infty
{Z_{lmk}\left[r(t),\iota(t)\right]\over\omega_{mk}^2}\,
S_{lm}(\vartheta)\,
\rme^{\left[\rmi(m\varphi - \omega_{mk}t)\right]}
\label{waveform}
\end{equation}
Here, $\mu$ is the mass of the small body, $D$ is the luminosity
distance to the source, $[r(t),\iota(t)]$ is the inspiral trajectory,
$Z_{lmk}[r,\iota]$ is a complex amplitude computed with the Teukolsky
equation, $\omega_{mk} = m\Omega_\phi + k\Omega_\theta$, $S_{lm}$ is a
spin-weighted spheroidal harmonic, and $(\vartheta,\varphi)$ is the
angular position of the observer relative to the spin axis.  Perhaps
the most effective demonstration of the characteristics of these
waveforms is given by converting the functions $h_+(t)$ and
$h_\times(t)$ into sounds; the reader is invited to visit the URL
given in {\cite{webwave} and listen to the sounds available there.
Some of the features one can hear in these waveforms are rather
surprising.  For instance, in several cases, the wave chirps {\it
down} as well as chirps up: a portion of the sound has decreasing
frequency.  Despite the many simplifications that were imposed in
order to compute these waves, they have a rather complex and ornate
character.

As the challenges discussed earlier are surmounted we will be able to
develop waveforms that incorporate even more structure and complexity.
The surprising features that were found for simple circular orbits
will doubtless be joined by more surprises, increasing the complexity
of the waveforms.  This complexity will make it difficult to detect
and analyse the waves in LISA data, but is indicative of how much can
be learned from their observation.

I would like to thank Pat Brady, Curt Cutler, Mike Hartl, Janna Levin,
Lee Lindblom, Sterl Phinney, Tom Prince and Kip Thorne for many useful
discussions and support.  This research was supported by NSF Grants
AST-9731698 and AST-9618537 and NASA Grants NAG5-6840 and NAG5-7034.

\end{document}